\newcommand{\solm}{M$_{\odot}$}
\newcommand{\rf}{\par\noindent\hangindent 15pt {}}

\title{The Galactic Center Black Hole Laboratory}

\author{\\
Eckart, A.$^{1,2}$\footnote{Email: eckart@ph1.uni-koeln.de},
Britzen, S.$^{2}$,
Valencia-S., M.$^{1}$, 
Straubmeier, C.$^{1}$,
\\
Zensus, J.A.$^{2,1}$
Karas, V.$^{3}$,
Kunneriath, D.$^{3}$,
Alberdi, A.$^{4}$,
\\
Sabha, N.$^{1}$,
Sch\"odel, R.$^{4}$,
Puetzfeld, D.$^{5}$ 
\\
\\
\\
1) I. Physikalisches Institut, Universit\"at zu K\"oln,\\
           Z\"ulpicher Str. 77, 50937 K\"oln, Germany
\\
2)            Max-Planck-Institut f\"ur Radioastronomie, \\
           Auf dem H\"ugel 69, 53121 Bonn, Germany
\\
3)       
          Astronomical Institute, Academy of Sciences, \\
        CZ-14131 Prague, Czech Republic
\\
4) Instituto de Astrof\'isica de Andaluc\'ia (CSIC), \\
Glorieta de la Astronom\'a s/n, 18008 Granada, Spain
\\
5) ZARM, University of Bremen, \\
Am Fallturm, 28359 Bremen, Germany
}

\date{\today}

\documentclass[11pt,english]{article}
\usepackage[]{graphicx}

\begin{document}
\maketitle

\begin{abstract}
The super-massive 4 million solar mass black hole Sagittarius~A* (SgrA*) shows 
flare emission from the millimeter to the X-ray domain. 
A detailed analysis of the infrared light curves allows us to 
address the accretion phenomenon in a statistical way.
The analysis shows that the near-infrared flare amplitudes 
are dominated by a single state power law, with the low states 
in SgrA* limited by confusion through the unresolved stellar background.
There are several dusty objects in the immediate vicinity  of SgrA*. 
The source G2/DSO is one of them. Its nature is unclear. 
It may be comparable to similar stellar dusty sources in the region or may 
consist predominantly of gas and dust.
In this case a particularly enhanced accretion activity onto SgrA* 
may be expected in the near future. Here the interpretation of recent
data and ongoing observations are discussed.
\end{abstract}

\normalsize
Sagittarius (SgrA*) at the center of our Galaxy is a highly variable
radio, near-infrared (NIR), and X-ray source which is associated with a
$ 4 \times 10^{6}${M$_{\odot}$} super-massive central black hole (SMBH).
SgrA* is the closest SMBH and can be taken as a paradigm for quiescent or very 
Low Luminosity Active Galactic Nuclei (LLAGN).
It has been shown that the strong polarized (up to several 10\%) infrared 
flux density excursions (often referred to as flares) from SgrA* show patterns 
of strong gravity as expected from in-spiraling material very close to 
the black hole's horizon (Zamaninasab et al. 2010, 2011).
As a consequence of the strong gravitational field of the black hole 
these patterns express themselves in characteristic variations of the light curve due to the
effect of relativistic boosting, light bending and the rotation of the polarization angle
(Zamaninasab et al. 2010, 2011, Eckart et al. 2006a, Broderick et al. 2005).
Therefore, it is mainly the polarization and the strong flux variability
that give us certainty that we study the immediate vicinity of a 
SMBH (Yoshikawa et al. 2013, Zamaninasab et al. 2010, 2011).

A fast moving infrared excess source G2, which is widely interpreted as a core-less gas 
and dust cloud (Gillessen et al. 2012), approaches SgrA* on a presumably elliptical orbit
covering the region of the high velocity S-stars close to SgrA*. 
The passage of this Dusty S-cluster Object (DSO) is expected to result in accretion
phenomena which will give an improved insight into the nature of the
immediate surroundings of the SMBH SgrA*.
The year 2013 certainly was only the first in an intense observing campaign during 
which the immediate vicinity of SgrA* is being monitored while the dusty object is close.
Recent orbit determinations expect the peri-apse passage to occur in April/May 2014
(Phifer et al. 2013).
Based on recent conference contributions (see Acknowledgments) on the Galactic Center 
this article concentrates on three major topics:
1) SgrA* and its environment,
2) its relation to extragalactic nuclei, and 
3) new instrumentation and perspectives.

\begin{figure}[!ht]
  \centering
  \includegraphics[width=0.99\textwidth]{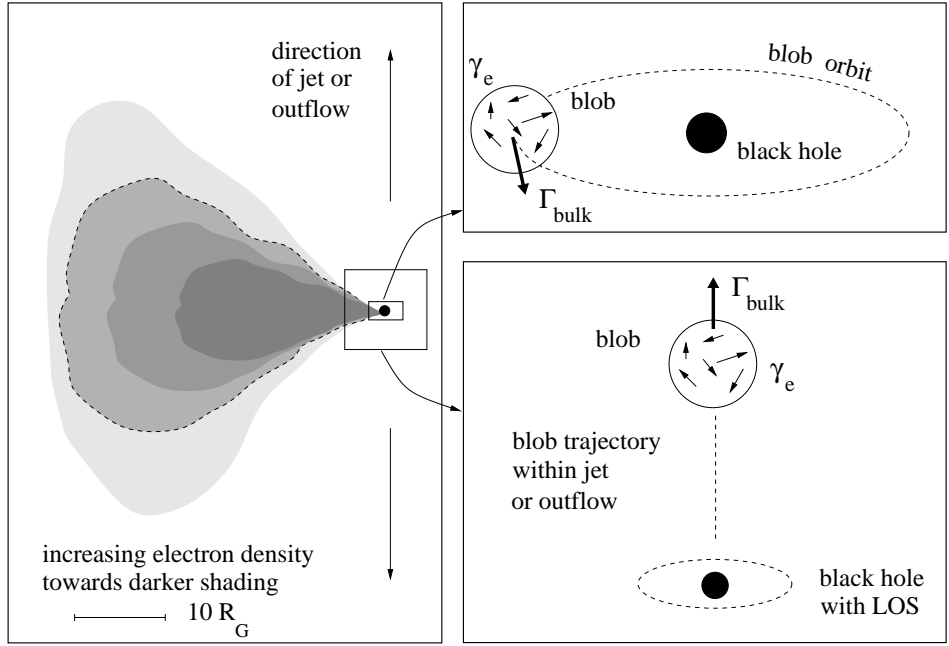}
  \caption{\small
Modeling approaches and the involved relativistic motions. We schematically compare the approaches through
disk (top right) and jet (bottom right) modeling to relativistic magneto hydrodynamic modeling (left).
}
  \label{Fig:models}    
\end{figure}

\section{The Galactic Center}
The DSO source approaching the super-massive black hole SgrA* at the center of the Milky Way 
has spawned great activities in observing that region covering the 
entire electromagnetic spectrum from radio, via infrared to X-ray wavelengths using 
telescopes across the world.
The upcoming events underline the importance of the Galactic Center as a laboratory to 
investigate and understand phenomena in the immediate environment of super massive black holes
(Eisenhauer 2010, Ghez 2009).
Gas and stars within the Galactic Center stellar cluster provide the fuel
for the central super-massive black hole and hence the reason for most of the
flux density variability observed from it.

\subsection{The variability of Sagittarius A*}
Progress has been made in the understanding of the emission process 
associated with the immediate surroundings of the super-massive black hole counterpart SgrA*  
as well as the three-dimensional dynamics and the population of the central stellar cluster.
There is also ample evidence of interactions between the cluster and SgrA*.

Series of monitoring observations in the near-infrared (NIR), X-ray, and sub-millimeter (sub-mm) regimes
accumulated over the years allowed us to perform for the first time detailed
statistical studies of the variability of SgrA*
(Witzel et al. 2012; Eckart et al. 2012).
The analyses show that the histogram of the near-infrared flux density is a
pure power-law and the emission process is most likely dominated by 
synchrotron radiation. 
In Witzel et al. (2012), we present a comprehensive data description for 
NIR measurements of SgrA*. 
We characterized the statistical properties of the variability of SgrA* in the near-infrared, which we found to be 
consistent with a single-state process forming a flux density power-law distribution. 
We discovered a linear rms-flux relation for the flux density range up to 12 mJy on a time scale of 24 minutes. 
This and the structure function of the flux density variations imply a phenomenological, formally nonlinear statistical 
behavior that can be modeled. In this way, we can simulate the observed variability and extrapolate its behavior to higher 
flux levels and longer time scales. 
SgrA* is also strongly variable in the X-ray domain
(Baganoff et al. 2003, Baganoff et al. 2001, Porquet et al. 2008, Porquet et al. 2003, Eckart et al. 2012 
and references there in, as well as, Nowak et al. 2012, Barriere et al. 2014 for a recent strong flares observed with
Chandra and NuSTAR).
The detailed statistical investigation by Witzel et al. (2012) also suggests that the past strong X-ray variations
that give rise to the observed X-ray echos can in fact be explained by the NIR variability histogram under the 
assumption of a Synchrotron Self Compton (SSC) process. 
SgrA* is extremely faint in the X-ray bands, 
though strong activity has been revealed through the detection of flares. 
Therefore, SgrA* is the ideal target to investigate the 
mass accretion and the ejection physics in the case of an extremely low 
accretion rate onto a super-massive black hole. This is actually the phase
in which super-massive black holes are thought to spend most of their lifetime.
The activity phase onset of a magnetar (see section below) at a separation of only 
about 3 arcseconds from the Galactic Center presented a problem 
for the SgrA* monitoring program in 2013
(Mori et al. 2013, Shannon et al. 2013, Rea et al. 2013).

Eckart, et al. (2012) present simultaneous observations and modeling 
of the millimeter (mm), NIR, and X-ray flare emission of SgrA*.
These data allowed us to investigate physical processes giving rise to the variable emission 
of SgrA* from the radio to the X-ray domain.
In the radio cm-regime SgrA* is hardy linearly polarized but shows a fractional circular polarization
of around 0.4\% (Bower, Falcke \& Backer 1999, Bower et al. 1999).
The circular polarization decreases towards the mm-domain (Bower 2003), where as
Macquart et al. (2006) report variable linear polarization from SgrA* of a few percent in the mm-wavelength domain.
The observations reveal flaring activity in all wavelength bands.
The polarization degree and angle in the sub-mm are likely linked to the magnetic field structure 
or the general orientation of the source.
In general - the NIR emission is leading the sub-mm with a delay of about one to two hours (see below)
and the excursions in the NIR and X-ray emission are rather simultaneous.
As a result we found that the observations can be modeled as the signal from an 
adiabatically expanding source component
(Eckart et al. 2008b, Yusef-Zadeh et al. 2006, Eckart et al. 2006b).
of relativistic electrons emitting via the synchrotron/SSC process.
A large fraction of the lower energy mm/cm- flux density excursions is not necessarily correlated 
with the NIR/X-ray variability
(e.g. Dexter \&  Fragile 2013, Dexter et al. 2013, and details and further references Eckart et al. 2012).
One may compute the SSC spectrum produced by up-scattering
of a power-law distribution of
sub-mm-wavelength photons into the NIR and X-ray domain by using the
formalism given by Marscher (1983) and Gould (1979).
Such a single SSC component model may be too simplistic,
although it is considered
as a possibility  in most of the recent modeling approaches.
It does not take into account possible deviations from the
overall spectral index of $\alpha$=1.3
at any specific wavelength domain like the NIR or X-ray regime.

The number density distribution of the relativistic electrons responsible for the 
synchrotron spectrum can be described by 
\begin{equation}\label{Equation:Gamma}
N(E) = N_0 E^{-2 \alpha +1}~~~,
\end{equation}
with  $\gamma_e$ between  $\gamma_1$ and  $\gamma_2$
which limit the lower and upper bound of the relativistic electron spectrum
\begin{equation}\label{Equation:spectrum}
\gamma_1 mc^2 < E= \gamma_e m c^2 <  \gamma_2 m c^2~~~.
\end{equation}
Lorentz factors $\gamma_e$ for the emitting electrons of the order of
a few thousand are required to produce a sufficient SSC flux in the
observed X-ray domain. In addition the relativistic bulk motion of 
the orbiting or outward traveling component
is described by the bulk  Lorentz factor $\Gamma$ (see Fig.~\ref{Fig:models}).
On the left of this figure we show a sketch of a typical relativistic electron 
density distribution resulting from MHD calculations
(e.g. Dexter et al. 2010,  Dexter \&  Fragile 2013, Moscibrodzka \& Falcke 2013). We show a cut through only 
one side of the three dimensional structure. The central plane and outflow region 
resulting from these calculations can be modeled in different, dedicated approaches (top and bottom right).

\begin{figure}[!ht]
  \centering
  \includegraphics[width=0.99\textwidth]{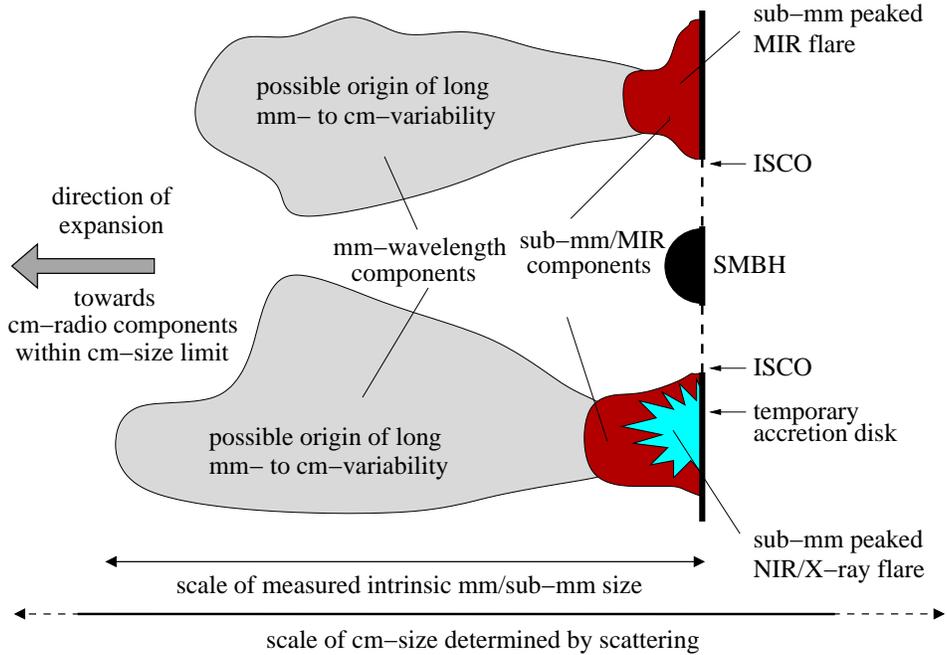}
  \caption{\small
Sketch of a possible source structure for the accretion disk around the SMBH
associated with SgrA* following Fig.12 in Eckart et al. (2008a).
}
  \label{Fig:diskmodel}    
\end{figure}

In the central plane modeling the flux variations are assumed to be the result of the motion of an orbiting blob
or a hot spot (e.g. Eckart et al. 2006a).
In the outflow model the flux variations are assumed to be due
to the ejection of a blob and its motion along the jet 
(with bulk motions close to the speed of light) or 
a much slower overall outflow component. 
In this case - for VLBI observations - the larger outflow extent at increasingly lower radio frequencies would be
hidden by the decreasingly lower angular resolutions due to interstellar scattering.
In Fig.\ref{Fig:diskmodel} the accretion disk (here assumed to be edge-on) is shown as a  vertical thick line 
to the right, the dashed part indicates the disk sections in the back- and foreground.
Extending to the left we show one side above the disk.
Here higher energy flare emission (lower part) is assumed to be responsible for the observed
NIR/X-ray flare emission.
Lower energy flare emission (upper part) may substantially contribute to long wavelength
infrared emission.
In addition to the expansion towards and beyond the the mm-source size, radial and azimuthal
expansion within the disk may occur.
Hence, long mm/cm-wavelength variability may originate from different source components of SgrA*
and may be difficult to be disentangled based on radio data alone.

In Fig.\ref{Fig:models}
the relativistic boosting vectors for the electrons $\gamma_e$ and the bulk motion $\Gamma_{bulk}$ are not 
drawn to a proper relative scale. One can assume $\gamma_e$ $\ge$ $\Gamma_{bulk}$.
In the case of relativistically orbiting gas as well as relativistic outflows
one may use modest values for $\Gamma$.
Both dynamical phenomena are likely to be relevant in the case of SgrA*.
The size of the central plane synchrotron component is assumed to be of the
order of or at most a few times the Schwarzschild radius.
From the overall variable radio/sub-mm spectrum spectrum we can assume a turnover frequency $\nu_m$ of a few 
100~GHz (see details in Eckart et al. 2012).
The motion of the synchrotron emitting cloud can be described via

\begin{equation}\label{Equation:delta}
\delta=\Gamma_{bulk}^{-1}(1-\beta cos \phi )^{-1}~~,
\end{equation}
\begin{equation}\label{Equation:Gamma}
\Gamma_{bulk}= (1-\beta^2)^{-1/2}~~.
\end{equation}
Here  $\beta= v/c $ and $v$ is the speed of the bulk motion of the synchrotron cloud,
$\delta$ is the Doppler factor and $\phi$ the angle to the line of sight.
For a relativistic bulk motion with $\Gamma_{bulk}$ around 
 1.7$\pm$0.3 (i.e. angles $\phi$ of about $30^{\circ} \pm 15^{\circ}$) the corresponding magnetic field
strengths are often assumed to be of the order of a few ten Gauss,
which is also within the range of magnetic fields expected for RIAF models
(e.g.  Yuan et al. 2006, Narayan et al. 1998).

\begin{figure}[!ht]
  \centering
  \includegraphics[width=0.99\textwidth]{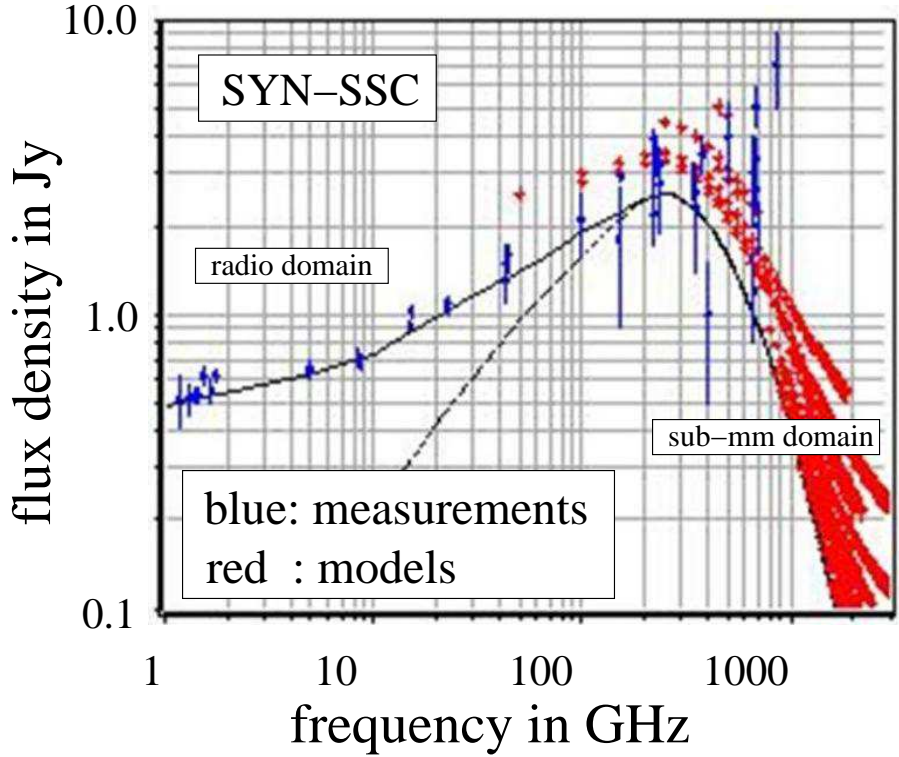}
  \caption{\small
The variable radio spectrum of SgrA*: Measurements and model results
(see text and Eckart et al. 2012 for details).
}
  \label{Fig:spectrum}    
\end{figure}

Modeling of the light curves shows that (at least for the brighter events) 
typically the sub-mm flux density excursions follow the NIR emission with a delay of 
about one to two hours with an expansion velocity of about 0.01c-0.001c
(Eckart et al. 2008b, Yusef-Zadeh et al. 2006, Eckart et al. 2006b).
We find source component sizes of around one Schwarzschild radius, flux densities of a few Janskys, 
and steep spectral indices. 
Typical model parameters suggest that either the adiabatically expanding source components have 
a bulk motion larger than its expansion velocity 
or the expanding material contributes to a corona or disk, 
confined to the immediate surroundings of SgrA*. 
For the bulk of the synchrotron and SSC models, we find synchrotron turnover frequencies in the 
range of 300-400 GHz. For the pure synchrotron models, this results in densities of relativistic 
particles in the mid-plane of the assumed accretion flow of the order of 10$^{6.5}$ cm$^{-3}$, 
and for the SSC models the median densities are about 
one order of magnitude higher. However, to obtain a realistic description of the frequency-dependent 
variability amplitude of SgrA*, models with higher turnover frequencies and even higher 
densities are required. 
This modeling approach also successfully reproduces the degree of flux density variability across 
the radio to far-infrared spectrum of SgrA*. 
In Fig.~\ref{Fig:spectrum} we show observed flux densities of SgrA* taken from the 
literature (blue) compared to a combined
model that consists of the fit given by Falcke et al. (2000), Marrone et al. (2008) (black line), 
and Dexter et al. (2010) (black dashed line). We plotted in red the spectra of synchrotron self-absorption frequencies for
the range of models. Here we show results for the preferred synchrotron plus SSC (SYN-SSC) model
that most closely represents the observed variability of SgrA*.

Valencia-S. et al. (2012) present theoretical polarimetric light curves expected in the 
case of optically thin NIR emission from over-dense regions close to the marginal stable orbit
(see also Broderick et al. 2005, Eckart et al. 2006a, Zamaninasab et al. 2010, 2011).
Using a numerical code the authors track the time evolution of detectable polarization 
properties produced by synchrotron emission of compact sources in the vicinity of the black hole. 
They show that the different setups lead to very special patterns in the time-profiles 
of polarized flux and the orientation of the polarization vector. As such, they may be used 
for determining the geometry of the accretion flow around SgrA* 
(see also Karas et al. 2011, Zamaninasab, et al. 2011).

During the 2013 Bad Honnef and the Granada conference (see Acknowledgments) 
efforts to monitor SgrA* during the DSO fly-by and first observational results from 2013 were reported by
Akiyama, et al. (2013ab), Eckart et al. (2013abc), Jalali et al. (2013), Meyer et al. (2013), Phifer et al. (2013).
The NRAO Karl G. Jansky Very Large Array (VLA) is undertaking an ongoing community 
service observing program to follow the expected encounter of the DSO 
cloud with the black hole SgrA* in 2013/14 (Chandler \& Sjouwerman 2013).
The NRAO VLA has been observing the Sgr~A region since 
October 2012 on roughly a bi-monthly interval, 
cycling through eight observing bands.
For monitoring the flux densities and in particular the radio spectral indices 
the short wavelength observations ($\lambda$$<$6cm) are most useful. 
For 2012/13 no particular flux density variation 
was detected that could be attributed to the interaction between SgrA* and the DSO. 
This may be linked with the fact that the newly determined periapse passage 
is now expected to happen in April/May of 2014 (Phifer et al. 2013), i.e. later than originally anticipated. 
However, in the radio-shock
frame in which variations of up to several Janskys were expected even during the pre-periapse time. 
Hence, the lack of strong radio flares 
indicates that the medium is less dense than expected and/or that the bow-shock size i.e. the
cross-section of the dust source is much smaller than assumed
(Narayan, et al. 2013, Crumley, et al. 2013, Sadowski, et al. 2013, Yusef-Zadeh, et al. 2013, Shcherbakov, et al. 2014).

\subsection{VLBI imaging of SgrA*}
There is also profound progress in imaging and modeling of the central putative accretion disk
of SgrA* as well as the jet that may be associated with the source
(Falcke \& Markoff, 2013, Moscibrodzka \& Falcke, 2013, Valencia-S, M. et al. 2012).
In fact imaging of SgrA* may turn out to be a Rosetta Stone in the attempts of distinguishing between 
different relativity theories of black holes
(e.g. Boller \&  M\"uller, 2013, on astronomical tests of general relativity and the
 pseudo-complex theory).

VLBI (Very Long Baseline Interferometry) observations at very short millimeter 
radio wavelengths can overcome the effects of interstellar scattering 
and allow us to study the source intrinsic structure of SgrA*.
Large mm/sub-mm facilities like the 
VLBA (Very Long Baseline Array), 
VERA (VLBI Exploration of Radio Astrometry), 
ALMA (Atacama Large Millimeter Array), 
PdBI (Plateau de Bure Interferometer) and the sensitive mm-telescopes in the
EVN (European VLBI Network) - such as the IRAM 30m and the 100m Effelsberg telescopes - are participating in this effort,
which will eventually culminate in the project EHT (Event Horizon Telescope), 
a VLBI array especially designed to image the structures close to the
event horizons of the larges SMBHs in the sky - namely SgrA* and M87,
with 1 Schwarzschild radius extending to an angular size of about 10$\mu$as and 3.7$\mu$as, respectively.
Multi-epoch imaging observations will allow to constrain the locations and sizes of the flaring region of SgrA* 
within the putative temporal accretion disk of the accretion stream/flow towards or an accretion wind from SgrA*.
These measurements will also constrain the acceleration processes (e.g. magnetic reconnection events or 
non-axisymmetric standing shocks) that give rise to the population of relativistic electrons and the variable 
emission we see from SgrA*.
Ultimatly, alternative black hole models will be probed and attempts to test the black hole no-hair theorem 
will be possible with the new VLB mm/sub-mm facilities
(Broderick, et al. 2014, Fish et al. 2014, Akiyama, et al. 2013ab, Huang, et al. 2012, Broderick, et al. 2011,
Broderick, et al.  2011, Fish et al. 2011, Lu, R.-S., et al.  2011).

These VLBI experiments will eventually enable spatially resolved studies on sub-horizon scales, leading 
to an unprecedented exploration of a putative predicted black-hole shadow 
(e.g. Huang et al. 2007)
as an evidence for light trapping by the black hole as well as its interaction with 
the surrounding material. 
It will be possible to monitor the possible expansion of source components 
during flare activity.
Furthermore, when a rotating black hole is
 immersed in a magnetic field of external origin, the gravito-magnetic 
interaction is capable of triggering the magnetic reconnection, 
accelerating the particles to very high energy 
(Karas et al. 2012, 2013; Morozova et al. 2014). 
This frame-dragging phenomenon is particularly interesting in the 
context of exploring the strong-gravity effects in astrophysical black holes 
because the effect does not have a Newtonian counterpart and it operates 
on the border of the ergospheric region (Koide \& Arai 2008), 
i.e. very close to the black hole horizon, and it can be probed with 
the future EHT. Also, one can investigate if the black hole proximity generates 
conditions favourable to incite the magnetic reconnection that eventually 
leads to plasma heating and particle acceleration. 
This effect could contribute to the flaring activity.

\subsection{The importance of dusty sources close to the center}
\begin{figure}[!ht]
  \centering
  \includegraphics[width=0.99\textwidth]{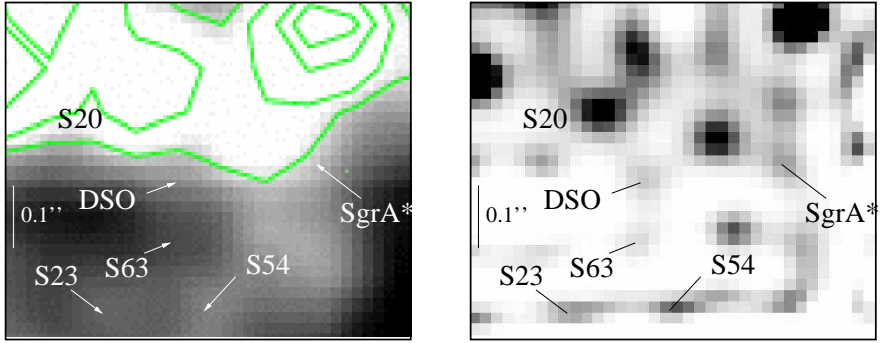}
  \caption{\small
The DSO detected in its K-band continuum emission in 2010 SINFONI data.
Left: The original image (positive greyscale); 
Right: A LUCY deconvolved image (negative greyscale) shown at an 
angular resolution close to the diffraction limit of the VLT UT4.
}
  \label{Fig:DSO}    
\end{figure}

A major discussion point is if and how the DSO source will be disrupted during
its peri-bothron\footnote{Peri- or apo-bothron is the term used for 
peri- or apoapsis - i.e. closest or furthest separation - for an elliptical orbit with a black hole present 
at one of the foci.  
As already mentioned by Frank and Rees (1976) word 'bothros' was apparently first suggested in the context of 
black holes by W.R. Stoeger.
It originates from the greek word 
\`o $\beta$\'o$\theta \rho o \varsigma$ 
with the equivalent meaning of 'the sink' or 'the deep dark pit'.
} passage.
It may be only its dusty envelope that will be disrupted
since the K$_s$-band identifications
of the source suggest that it can also be associated with a star (Eckart et al. 2013a).
In addition to the VLT NACO and the Keck NIRC detections of the 
DSO NIR continuum emission (Eckart et al. 2013bc),
here we show the detection of the DSO continuum at about K$\sim$19 using SINFONI data (Fig.~\ref{Fig:DSO}).
The detection of the continuum emission in data sets taken with three different 
instrumental setups over many years strengthens the case for a substantial
continuum emission from that dusty source.
As posted in the astronomer's telegram No.6110 on 2 May 2014 (Ghez et al. 2014), the DSO was detected 
 3.8$\mu$m during its peri-bothron passage around the central black hole SgrA*. Hence, it appears to be intact
and up to this point not yet heavily affected by tidal effects. 
This clearly supports our finding (Eckart et al. 2013bc) that it may very well be a dusty star
rather than a pure gas and dust cloud.

In contrast to a pure dust and gas nature of the DSO its possible 
stellar (i.e. a dust enshrouded star) nature is discussed and partially favoured in
Meyer et al. (2013, 2014), Eckart et al. (2013abc), Scoville \& Burkert (2013)
Ballone et al. (2013), Phifer et al. (2013).
Eckart et al. (2013a) investigate the possible mass transfer across 
Lagrange point L1 in a simple Roche model.
If the star has a mass of about 1\solm,  the separation of $L1$ from it will be
about 0.1~AU. For a Herbig Ae/Be stars with 2-8~\solm ~that distance will be between
0.2 and 0.5~AU. For a typical S-cluster stellar mass of $\sim$20-30~\solm ~the separation 
will be closer to one AU. 
The interferometrically determined  inner ring sizes that one typically finds for young Herbig Ae/Be and
T~Tauri stars can indeed be as small as 0.1-1~AU (Monnier \& Millan-Gabet 2002).
Any stellar disk or shell may already have been stripped 
substantially if the DSO has performed more than a single orbit.
If the source has a size of about 1~AU 
(as determined from its MIR-luminosity; Gillessen et al. 2012a) 
then a significant amount of the dusty circumstellar material
may pass beyond $L1$ during peri-bothron passage. 
This material will then start to move into the Roche lobe associated with SgrA*.

However, it is not at all clear what will happen to the transferred material  after 
the peri-bothron passage around May 2014 or beyond. 
The fact that this dusty object may be a dust enshrouded star rather than a dust 
cloud will have an influence on the expected flux density variations resulting 
from the close approach. They may be much weaker than expected.
Simulations (e.g.  Burkert et al. 2012, Schartmann et al. 2012,
see also Zajacek, Karas \& Eckart 2014) that
have discussed the feeding rate of SgrA* as a function of radius
indicate that a portion of the material may fall towards SgrA*.
If SgrA* is associated with a significant wind on scale of the peri-bothron separation, 
then a large part of the material may be blown away again by an out-bound accretion wind.
Shcherbakov \& Baganoff (2010) have discussed the feeding rate of SgrA* as a function of radius.
Based on their modeling one may suggest that the bow-shock sources X3 and X7 (Muzic et al. 2010) are  
still in the regime in which most of the in-flowing mass is blown away again.
Another case for comparison is the star S2. During its peri-bothron passage the star has been well 
within the zone in which matter of its (weak) stellar wind could have been accreted by SgrA*. 
The DSO peri-bothron will be at a larger radius than that of S2 (Phifer et al. 2013). 
This may imply that 
no enhanced accretion effect will result from it during the peri-bothron passage. 
Until May 2014 no increase in variability and no significant flux density increase well above normal levels 
has been reported in the radio to X-ray domains.

The fate of the DSO and the cometary sources X3 and X7 underline the importance
of investigating the wind properties in the vicinity of SgrA* in more detail.
IRS~8 is a unique possibility to study the bow shock properties and polarization 
features in the dusty environment at the Galactic Center.
Based on a detailed study of near-infrared emission 
Rauch et al. (2013) present interstellar dust properties for 
the northern arm in the vicinity of the IRS~8 bow shock. 
This study allowed us for the first time to determine the relative positioning of 
IRS~8 with respect to the northern arm and the 
super-massive black hole SgrA*. The result indicates that the
central star of IRS~8 is in fact located closer towards the observer than the northern arm.
In Eckart et al. (2013a) we investigated the near-infrared 
proper motions and spectra of infrared excess sources at the Galactic Center.  
The work concentrated on a small but dense cluster of comoving 
sources (IRS13N) located ~3'' west of SgrA*.  Our analysis shows that 
these stars are spectroscopically and dynamically young and can indeed be 
identified with continuum emission at 2 microns and shortward, indicating that these mid-infrared 
sources are not only dust sources but young stars.
The possibility of ongoing star formation at the Galactic Center is 
supported through simulations by Jalali et al. 2014 (submitted) and  Jalali et al. (2013).
In fact the DSO may be a representative of dusty sources similar to those
discussed in Eckart et al. (2006b; see their Fig.14;
compare also to the discussion of sources X3 and X7 given by Muzic et al. 2010).
Meyer et al. (2014) present NIR spectroscopic data of several of these sources.
They also show that the DOS does not seem to be unique, since several red emission-line objects 
can be found in the central arcsecond. 
In summary, Meyer et al. (2014) conclude that it seems more likely that G2 is ultimately a
stellar source that is clearly associated with gas and dust (see also Eckart et al. 2013abc).


\subsection{Stellar dynamics and tests of relativity}
SgrA*, the super-massive black hole at the center of the Milky Way, 
is surrounded by a small cluster of high velocity stars, known as the S-stars
(Eckart \& Genzel 1997).
Sabha et al. (2012) aimed at constraining  the amount and nature of the stellar 
and dark mass that is associated with the cluster in the immediate vicinity of SgrA*. 
The authors use near-infrared imaging to determine the Ks-band luminosity function 
of the S-star cluster members, the distribution of the diffuse background 
emission and the stellar number density counts around the central black hole. 
This allows us to determine the stellar light and mass contribution expected 
from the faint members of the cluster. 
Sabha et al. (2012) then use post-Newtonian N-body simulations to investigate the effect of stellar 
perturbations on the motion of S2, as a means of detecting the number and masses 
of the perturbers. The authors find that the stellar mass derived from the Ks-band 
luminosity extrapolation is much smaller than the amount of mass that might 
be present considering the uncertainties in the orbital motion of the star S2. 
Also the amount of light from the fainter S-cluster members is below the 
amount of the residual light at the position of the S-star cluster after one removes 
the bright cluster members. If the distribution of stars and stellar remnants 
is peaked near SgrA* strongly enough, observed changes in the orbital 
elements of S2 can be used to constrain both the masses and the number of objects inside its orbit. 
Based on simulations of the cluster of high velocity stars we find that in
the NIR K-band - close to the confusion level for 8 m class telescopes -
blended stars will occur preferentially near the position of SgrA*
which is the direction towards which we find the highest stellar density.
These blended stars consist of several faint, (with the current facilities) 
individually undetectable stars that get aligned along the line-of-sight, 
producing the visual effect of a new point source. 
The proper motion of stars and the corresponding velocity dispersion
leads to the fact that such a blended star configuration dissolves typically after 3 years.

Stars that get very close to the super massive black hole are ideal probes to
analyse the gravitational field and to search for effects of relativity due 
to the presence of the high mass concentration and its effect on space time.
This can be done by tracing the orbit of stars through proper motions and radial velocities.
As discussed in Zucker et al. (2006) relativistic effects should express themselves 
spectroscopically. The redshift $z$ of a black hole orbiting star can be written as:
\begin{equation}\label{eq:aa1}
z = \Delta\lambda/\lambda = B_0 + B_1\beta + B_2\beta^2 + O(\beta^3)
\end{equation}
with $B_0$ being an offset,
$B_1\beta$ describing the Doppler velocity  
and $B_2\beta^2$ expressing the relativistic effects.
Here the value $B_2$ contains equal contributions from the gravitational redshift and the 
special relativistic transverse Doppler effect.
The combined effect gives a redshift that is about an order of magnitude larger than
the currently achieved spectral resolution of $\delta\lambda/\lambda$$\sim$10$^{-4}$.
For S2 one expects about a 150-200 km/s signal measurable over a few months 
on top of an orbit-depending radial velocity of more than 4000 km/s.
Expectations are high that this will be observable during the next peri-bothron
for S2 around 2017.9$\pm$0.35 (Gillessen et al. 2009b, Eisenhauer et al. 2003) 
or S2-102 around 2021.0$\pm$0.3 (Meyer et al. 2013).
Realistically, however, one needs several stars  on different orbits 
to detect the relativistic effect with certainty (Zucker et al. 2006; see also  
Rubilar \&  Eckart 2001 for peri-bothron shift).
Alternatively, one has to find stars that are (or get) closer than S2 
and S2-102 (Meyer et al. 2013; see below) to SgrA*.

Detailed imaging and the analysis of proper motions may be another way to trace relativistic effects.
An important deviation from Keplerian motion occurs as a result of
relativistic corrections to the equations of motion, which to the lowest
order predict a certain advance of the argument of 
peri-bothron each orbital period.
Choosing $a = 5.0$~mpc and $e= 0.88$ for the semi-major axis and eccentricity 
of S2, respectively, and assuming a black hole mass of $M_\bullet=4.0\times 10^6 $~M$_\odot$ this advance will be
\begin{equation}\label{Equation:DomegaGR}
\left(\Delta\omega\right)_\mathrm{GR} 
= \frac{6\pi GM_\bullet}{c^2a(1-e^2)} \approx 10.8^\prime.
\end{equation}
The relativistic precession is prograde, and leaves the orientation of the orbital
plane unchanged.

The location of the peri-bothron advances for each orbital period
due to the spherically-symmetric component of the distributed mass that 
is resolved by the elliptical orbit of the star.
The amplitude of this Newtonian ``mass precession'' is
\begin{equation}\label{eq:nuM}
\left(\Delta\omega\right)_\mathrm{M} = -2\pi G_\mathrm{M}(e,\gamma)\sqrt{1-e^2}\left[\frac{M_\star}{M_\bullet}\right].
\end{equation}
Here, 
$M_\star =  M_\star(r<a)$
is the distributed mass within a radius $r=a$, and $G_\mathrm{M}$ is
a dimensionless factor of order unity that depends on $e$ and on the power-law index
of the density, $\rho\propto r^{-\gamma}$ (Merritt 2012).
In the special case $\gamma=2$,
\begin{equation}
G_\mathrm{M} = \left(1+\sqrt{1-e^2}\right)^{-1}
\approx 0.68\ \mathrm{for\ S2}~~,
\end{equation}
so that
\begin{equation}\label{Equation:DomegaM}
\left(\Delta\omega\right)_\mathrm{M} \approx 
-1.0^\prime\; \left[\frac{M_\star}{10^3 \mathrm{~M}_\odot}\right].
\end{equation}
Mass precession is retrograde, i.e., opposite in sense to the relativistic precession.

The contribution of relativity to the peri-bothron advance is determined 
uniquely by the known values of $a$ and $e$.
A measured $\Delta\omega$ can then be used to constrain the 
mass enclosed within S2's orbit, by subtracting
$(\Delta\omega)_\mathrm{GR}$ and comparing the result with Equation~(\ref{Equation:DomegaM}).
So far, this technique has yielded only upper limits on $M_\star$ of 
$\sim 10^{-2}M_\bullet$ (Gillessen et al. 2009a).
First robust upper limits of that order have been derived by Mouawad et al. (2003, 2005).
The role of relativistic effects versus the gravitational effects of the nuclear star cluster 
and a ring of stars are summarized by Subr et al. (2004).


  \begin{figure}[!ht]
  \centering
  \includegraphics[width=0.99\textwidth]{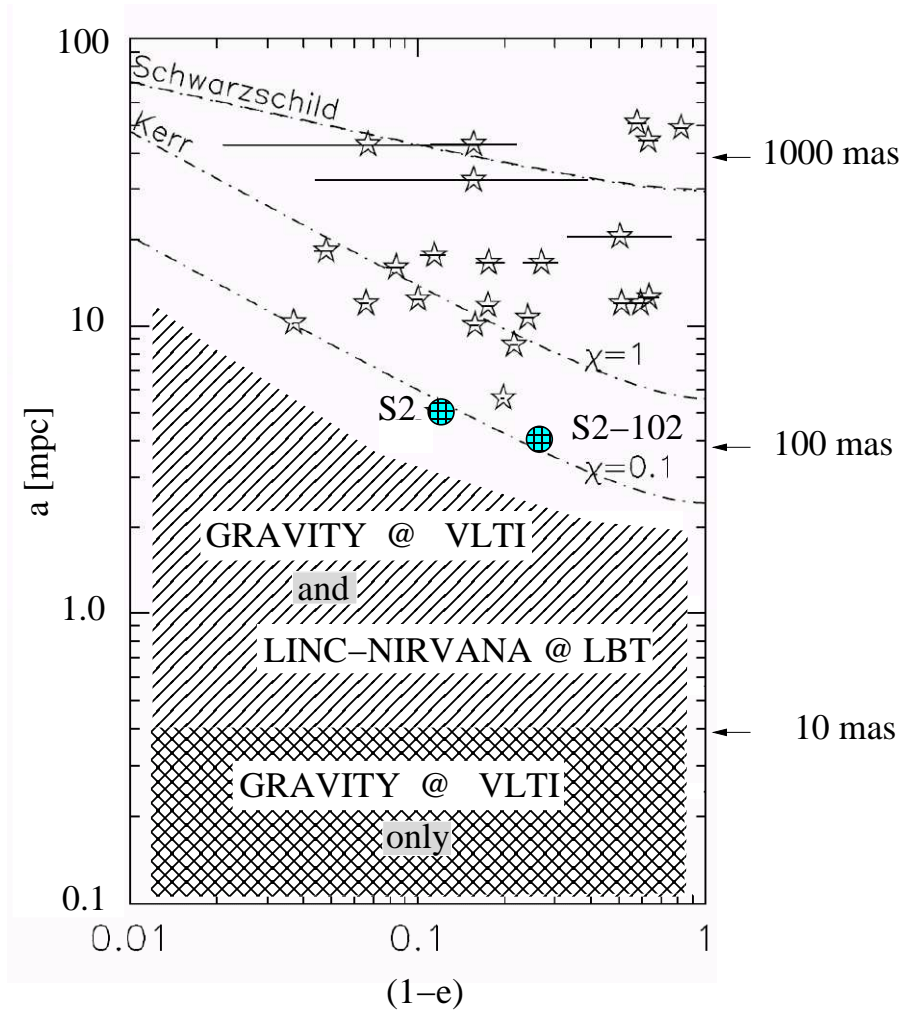}
  \caption{\small
An example for the location of the Galactic Center S-stars on the $(a,e)$-plane.
See details in the text and Antonini \& Merritt (2013), in particular the left panel
of their Fig.1.
}
  \label{Fig:stellarorbits}    
\end{figure}

\begin{figure}[!ht]
  \centering
  \includegraphics[width=0.99\textwidth]{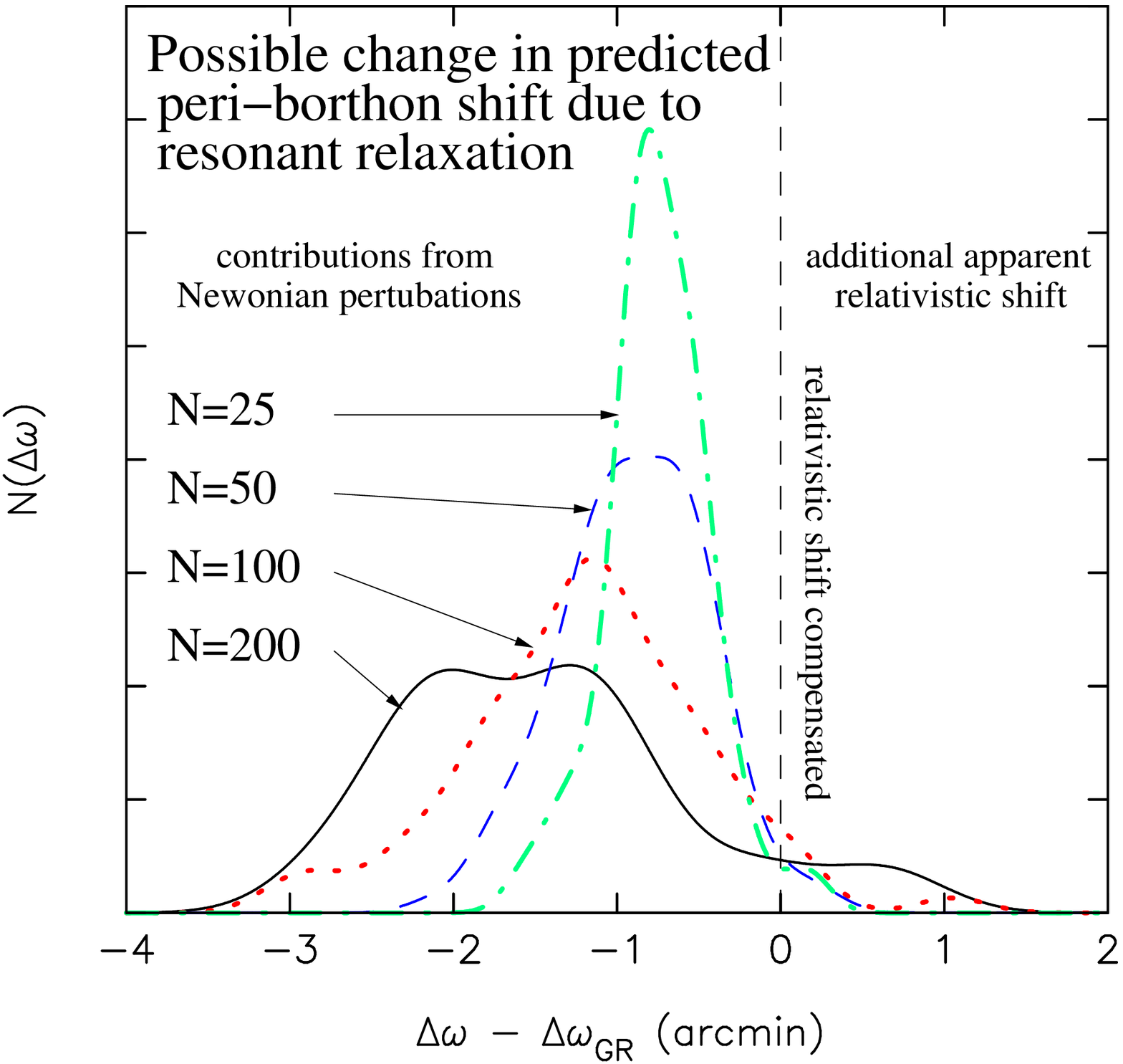}
  \caption{\small
Histograms of the predicted change in S2's argument of peri-bothron, $\omega$,
over the course of one orbital period ($\sim15.6$~yr; see also Sabha et al. 2012).
See text for a detailed description of the figure.
}
  \label{Fig:S2histogram}    
\end{figure}


In addition to the relativistic effects on the orbit of a test star, and the precession due to the smooth matter distribution, the granularity of the cluster stars and stellar remnants needs to be considered. The granularity of the distributed mass makes itself felt via the phenomenon of “resonant relaxation” (RR; Rauch \& Tremaine 1996, Hopman \& Alexander 2006). On current observational time scales the stellar orbits near SgrA* remain nearly fixed in their orientations. The perturbing effect of each field star on the motion of a test star (e.g. S2) can be approximated as a torque that is fixed in time and proportional to the mass $m$ of the field star. 
Sabha et al. (2012) show that the effects of RR are competing with the relativistic and Newtonian periastron effects on the orbits of the known high velocity S-cluster stars. 
Following Sabha et al. (2012), we show in Fig.~\ref{Fig:S2histogram}    
the predicted change in S2's orbital elements over the course of one orbital period ($\sim 15.6$ yr). The shift due to relativity, $\Delta\omega_\mathrm{GR} \approx 11^\prime$,
has been subtracted from the total; what remains is due to Newtonian perturbations from the field stars. Each histogram was constructed from integrations of 100 random realizations of the same initial model, with field-star mass $m = 10M_\odot$, 
and four different values of the total number: $N = 200$ (solid/black); 
$N = 100$ (dotted/red); $N = 50$ (dashed/blue); and $N = 25$ (dot-dashed/green). 
The average value of the peri-bothron shift increases with increasing $Nm$, as predicted by Equation (10). There is also a separate contribution that scales approximately as $\sim 1/\sqrt{N}$ and that results in a variance about the mean value. 

Sabha et al. (2012) show that the net effect of the torques from $N$ field stars is to change the angular momentum, $\mathbf{L}$, of S2's orbit according to
\begin{equation}\label{Equation:DeltaLRR}
\frac{|\Delta\mathbf L|}{L_c} \approx K\sqrt{N}\frac{m}{M_\bullet}\frac{\Delta t}{P}
\end{equation}
where $L_c$ is the angular momentum of a circular orbit having the same semi-major
axis as that of the test star.
Here Equation \ref{Equation:DeltaLRR} describes ``coherent resonant relaxation''.
Sabha et al. (2012) state that the normalizing factor $K$ should be of order unity (Eilon et al. 2009).

Changes in $\mathbf{L}$ imply changes in both the eccentricity of the test star orbit, 
as well as changes in its orbital plane.
Changes in the orbital plane can be described in a coordinate-independent way via the
angle $\Delta\theta$, where
\begin{equation}
\cos\left(\Delta\theta\right) = \frac{\mathbf{L}_1\cdot\mathbf{L}_2}{L_1L_2}
\end{equation}
and $\{\mathbf{L}_1, \mathbf{L}_2\}$ are the values of $\mathbf{L}$ at two times
separated by $\Delta t$ (Sabha et al. 2012).
For small values of $\Delta \theta$ this can be written as
\begin{equation}
\Delta\theta \approx \sqrt{2 - 2\frac{\mathbf{L}_1\cdot\mathbf{L}_2}{L_1L_2}} \approx
1 - \frac{\mathbf{L}_1\cdot\mathbf{L}_2}{L_1L_2} .
\end{equation}
If we set $\Delta t$ equal to the orbital period of the test star, the changes
in its orbital elements due to resonant relaxation are expected to be
\begin{eqnarray}
\left|\Delta e\right|_\mathrm{RR} &\approx& K_e \sqrt{N} \frac{m}{M_\bullet} , \\
\left(\Delta \theta\right)_\mathrm{RR} &\approx& 2\pi K_t\sqrt{N} \frac{m}{M_\bullet} ,
\end{eqnarray}
where $N$ is the number of stars having $a$-values similar to, or less than, that of
the test star (Sabha et al. 2012).
The constants $K_e$ and $K_t$ may depend on the properties of the orbits of the
field star distribution in the Galactic Center stellar cluster.
More details are given in Sabha et al. (2012).
The Kozai mechanism is another kind of resonant process that is thought to operate in the Galactic Center environment (Subr \& Karas 2005, Chang 2009, Chen \& Amaro- Seoane 2014).

Figure Fig.~\ref{Fig:stellarorbits}    (following Antonini \& Merritt 2013)
plots the location of the S stars in the $(a,e)$-plane (semimajor axis - ellipticity).
The data for the two currently closest known stars S2 and S2-102 
are indicated by pattern filled circles. 
For the two currently closest sources one finds: 
\begin{eqnarray}
\mathrm{S0-102}:  T_\mathrm{orbit} &=& 11.5\pm0.3\; \mathrm{yr} \ \ e=0.680\pm0.020 \ \  a=0.100''\pm0.010''\; \nonumber \\
\mathrm{S2}: T_\mathrm{orbit} &=& 15.6\pm 0.4\; \mathrm{yr} \ \  e = 0.883\pm0.003 \ \  a = 0.125''\pm 0.002''\;  \nonumber
\end{eqnarray}
(Meyer et al. 2013, Gillessen et al. 2009b, Eisenhauer et al. 2003; 
1'' at the distance of the Galactic Center corresponds to  alinear scale of about $39~mpc = 0.127~lyr = 8044~AU$).

Also plotted in Fig.~\ref{Fig:stellarorbits} are curves indicating where the effects of RR begin to be
mediated by relativistic precession of the ``test'' star.
The upper curve is the ``Schwarzschild barrier'' (Merritt et al. 2011); 
stars below this curve precess due to GR so rapidly that their precession,
rather than the mean precession rate of the field stars, determines the coherence
time over which the RR torques can act.
This is reflected in the horizontal tick marks which give the expected amplitude of eccentricity changes as an orbit precesses in the essentially fixed torquing field due to the field stars.
The location of the Schwarzschild barrier in the $(a,e)$-plane
depends somewhat on the assumed
spatial distribution of the stars in the nuclear cluster, but for most reasonable
distributions, S2 lies below the barrier (Antonini \& Merritt 2013).
Closer to SgrA*, frame-dragging torques due to the spin of SgrA* begin to make
themselves felt. 
The two lower curves on Fig. 6 mark where these torques begin to compete with
RR torques.
In this regime, orbits precess so rapidly that their eccentricities are essentially
unaffected by the $\sqrt{N}$ torques, but their orbital planes can still change
(``vector RR'').
Below the curves marked ``Kerr'' in 
Fig.~\ref{Fig:stellarorbits} are curves indicating where the effects of RR begin to be
changes in orbital planes due
to frame dragging dominate the changes due to vector RR (Merritt et al. 2010).
Stars in this regime can be used to test theories of gravity, e.g., ``no-hair''
theorems.

Fig.~\ref{Fig:stellarorbits} 
also shows the $(a,e)$-regions that will be experimentally accessible with the new upcoming interferometric instrumentation in the NIR like GRAVITY at the VLTI and LINC-NIRVANA at the LBT (see below). 
Only for stars at much smaller orbital separations from SgrA* than S2 or S2-102, 
as they may be found with infrared interferometers like GRAVITY at the VLTI or LINC-NIRVANA at the LBT (see below), are Newtonian perturbations (mass precession,
RR) negligible compared with the effects of GR.
However, even then one will need more than one star with different orbital elements in order to clearly 
demonstrate that the eventually observed orbital changes are truly due to relativistic effects 
(see e.g.  Merritt et al. 2010, Will 2008 , Zucker et al. 2006, Rubilar \& Eckart 2001).

\subsection{Effects due to stellar collisions}
The stellar density close to center is in excess of 10$^5$ \solm ~~per cubic-parsec.
This implies that in this region the evolution of stars is influenced by stellar collisions
(e.g.  Davies et al. 2011, Church et al. 2009).
Observations of the stars 
at the Galactic Center show that there is a lack of red giants 
within about 0.5~pc of the super-massive black hole. 
The very high stellar number densities this close to the SMBH 
imply that the giants -- or their progenitors -- may have been 
destroyed by stellar collisions. 
The derivation of collisional rates between different 
types of stars at the Galactic Center and their likely effects
are currently a field of intense investigation
in order to quantify how large a contribution stellar 
collisions could make to the puzzle of the missing red giants
(Dale et al. 2009).
As an example see recent contributions on star formation (see also references below), 
and on the structure of the 
central stellar cluster and stellar interactions in the Galactic Center area by
Sabha et al. (2012), Perets et al. (2014), Mastrobuono-Battisti \&  Perets (2013),
Haas \& Subr (2012), Subr \& Haas (2012), Haas, Subr \& Vokrouhlicky (2011), Feldmeier et al. (2013).

\subsection{Pulsars at the Galactic Center}
Another very promising way of investigating the super-massive black hole properties in 
great detail is to find pulsars orbiting SgrA*. 
If close enough, they
would allow us to measure the spin and quadrupole moment of the 
central black hole with superb precision
enabling us to test different theories of gravity
(Psaltis et al. 2012).
The number of pulsars in the central cluster will depend on the 
star formation history in the overall region.
The discovery of radio pulsations from a magnetar PSR~J1745-2900 with the 
Effelsberg telescope has highlighted the great value and the efforts 
that are currently being undertaken to find pulsars in the Galactic Center region
(Spitler et al. 2014, Lee et al. 2013, Mori et al. 2013, Eatough et al. 2013a, Eatough et al. 2013b).

\subsection{Star formation and the Galactic Center}
Star formation activity in the Galactic Center is an ongoing research topic
(e.g. Yusef-Zadeh et al. 2009, 2010, 2013).
Given the deep gravitational potential towards the very center,
stars cannot form in the same way they do throughout the Milky Way.
However, infrared observations have shown evidence that massive 
stars were formed in the hostile environment of SgrA* a 
few million years ago. 
VLA, ALMA and CARMA measurements suggest that 
star formation is still taking place in this region and has been going on during the
past few 10$^5$ years. A broad variety of possible star formation scenarios
can be discussed. Massive stars may have formed within a disk of 
molecular gas resulting from a passage of a giant molecular
cloud interacting with SgrA* and then dispersed after having formed these stars
(Nayakshin, Cuadra \& Springel 2007).
Dense clumps originating from the CND
loosing angular momentum and falling towards the very center may also
be a source of constant or episodic star formation in the Galactic Center 
(Jalali et al. 2013).
A fraction of the material associated with these phenomena  
may have accreted onto SgrA* probably in an episodic or transient manner (Czerny et al. 2013). 
This process may in part be responsible for the origin of the 
$\gamma$-ray emitting Fermi bubbles (Su,  Slatyer \&  Finkbeiner 2010).

\subsection{The Galactic Center on larger scales}

Radio polarization observations by the Parkes radio telescope in Australia have recently 
led to the discovery of giant radio lobes emanating from the Galactic 
nucleus (Zubovas \&  Nayakshin 2012, Bland-Hawthorn et al. 2013).
These lobes are largely coincident with the Fermi Bubbles discovered in the
$\gamma$-ray domain (Su,  Slatyer \&  Finkbeiner 2010).
However, the radio outflows extend to even larger angular scales, 
covering about 55 to 60 degrees north and south of the Galactic plane. 
It is likely that the radio lobes -- and the Fermi Bubbles -- are 
the result of the concentrated star formation occurring in the 
Central Molecular Zone of the Milky Way rather than signatures 
of putative activity of the super-massive black hole. 
These Fermi bubbles may be related to 
the giant magnetized outflows from the Galactic Center 
(Crocker 2012, Crocker et al. 2011, Crocker \& Aharonian 2011).
These phenomena link the Galactic Center nuclei of nearby active galaxies and 
larger scale events (e.g. amounts of gas escaping from the potential well of the galaxy).
Massive inflow of gas towards the centers, as well as
star formation or jet driven outflows from the nuclear regions, are frequently observed 
and are directly or indirectly linked with the accretion processes onto super-massive black holes.

\section{Extragalactic Nuclei}
The observational results obtained on the Galactic Center during the past 
few years need to be put into perspective. 
The nucleus of our Galaxy has an extremely low luminosity.  
However, it also demonstrates that violent activity may take place at times. 
Hence, a comparison between the center of the Milky Way and the nuclei of galaxies hosting 
Low Luminosity Active Galactic Nuclei (LLAGN) appears to be imperative.

Footprints of AGN feeding and feedback in LLAGN can be observed in many cases
(Garcia-Burillo \& Combes 2012, Garcia-Burillo et al. 2012, Marquez \& Masegosa 2008).
The study of the content, distribution and kinematics of interstellar 
gas is a key to understanding the fueling of AGN and star formation activity 
in galaxy disks. Current mm-interferometers provide a sharp view of the 
distribution and kinematics of molecular gas in the circumnuclear disks 
of galaxies through extensive mapping of molecular line (mainly CO and to some extent
high density tracers like HCN, HCO$^+$, CS etc). 
The use of molecular tracers specific to the dense gas phase 
can probe the influence of feedback on the chemistry and energy 
balance in the interstellar medium of galaxies. 
Radiative and mechanical feedback are often used as a mechanism 
of self-regulation in galaxy evolution 
as well as thermal and non-thermal AGN activity.
In the disks of galaxies the evolution is predominantly expressed in star formation and 
the evolution of stellar populations and the ISM properties.
In galactic nuclei the thermal part is represented mainly by the properties of the nuclear ISM
and by the NLR and BLR regions while the non-thermal part is dominated by 
synchrotron/SSC emission due to SMBH accretion and the presence of jets.

High angular resolution ($<$1'') observations with sub(mm)-interferometers 
(IRAM PdBI, ALMA) in the context of the NUclei of GAlaxies (NUGA) survey 
(e.g. Garcia-Burillo \& Combes 2012, Garcia-Burillo et al. 2012, 
Garcia-Burillo et al. 2003, Krips et al. 2007, Moser et al. 2013,
and upcoming more ALMA NUGA papers)
allow us to study the mechanisms responsible for fueling AGN and star formation 
activity in the central R$<$1 kpc disks of a 
sample of 25 active galaxies at the 10-100 pc scale. 
This study has revealed streaming motions towards and away from the nuclear 
region that do not necessarily have to be co-phased with current AGN activity.
In several cases these observations also reveal the presence of 
molecular circumnuclear disks and massive molecular outflows.

These phenomena are directly coupled to black-hole fueling, feedback and 
duty cycles that are essential for understanding the activity in galactic nuclei
(Mundell et al. 2003, 2004, 2007, 2009;
see also Combes et al. 2014, Heckman \& Best 2014).
The distribution of 
gas and stars in nearby galaxies traced by 3-D studies of molecular, 
neutral and ionized gas provides a unique view of the role of the 
multi-phase medium in triggering and fueling nuclear activity in 
galactic nuclei on scales ever closer to the central black hole. 
Radio- and mm-interferometers and in particular 3-dimensional imaging 
devices in the infrared and optical domain (like integral field units) 
allow us to obtain spectroscopic and photometric information at every position
of the field-of-view. Such observations have been used to elaborate
comparative studies of gaseous 
and stellar dynamics in active and quiescent galaxies. 
Studying the inner kiloparsec of AGN is particularly important, since 
the activity and dynamical time scales become comparable in this region.
These investigations also show that the variability observed 
in nearby radio-quiet AGN may provide 
parallels to our own Galactic Center. 
They also provide evidence that AGN duty cycles 
may be shorter than previously thought.


\begin{figure}[!ht]
  \centering
   \includegraphics[width=0.99\textwidth]{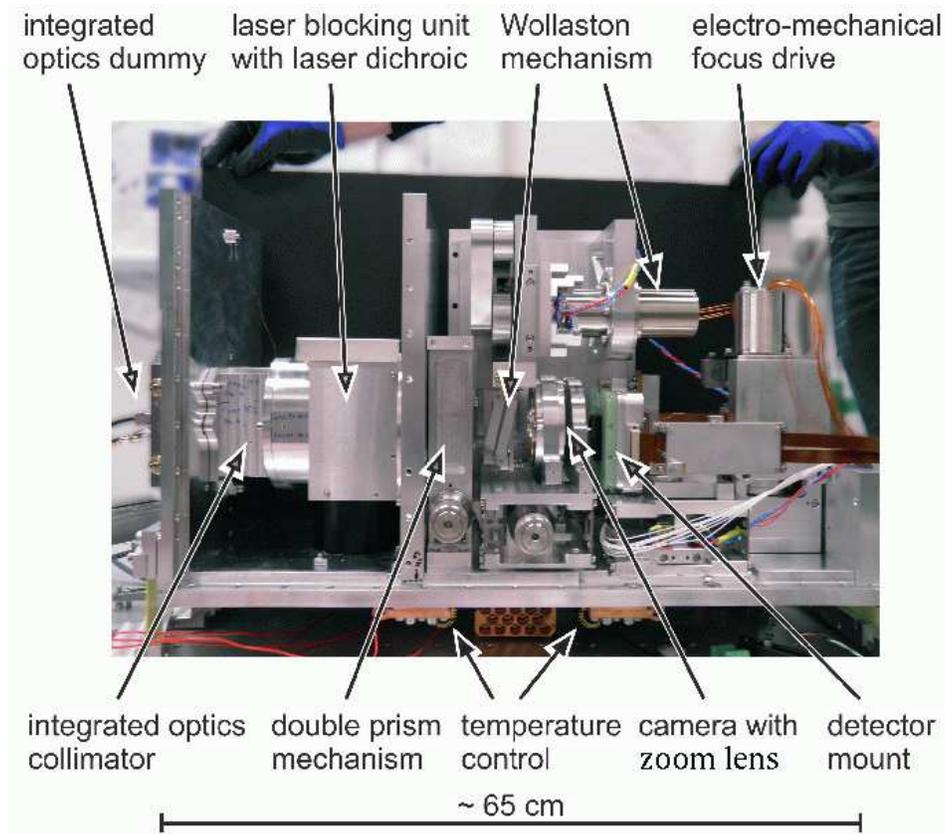}
  \caption{\small
The fringe tracking spectrometer of GRAVITY being under 
construction at the University of Cologne (Straubmeier et al. 2012).
}
  \label{Fig:Gravity}    
\end{figure}

\section{Black Hole Laboratory: New instrumentation and perspective}
An improved link between theory and observations will be provided by new and upcoming 
instrumentation such as the future capability of measuring polarization in the X-ray 
domain (Soffitta et al. 2012, 2013),
and direct VLBI imaging of the event horizon region for e.g. M87 as a representative of the largest
super massive black holes, and SgrA* as a representative for low-mass SMBHs 
(Dexter et al. 2012, Fish et al. 2014, Broderick et al. 2014).
The SKA (Square Kilometer Array) will be key instrument to stablish
the census of pulsars at the Galactic Center, to determine the star
formation history, and to provide the required precision to determine
the spin and quadrupole of the central BH, testing the theories
of gravity (e.g. Aharonian et al. 2013).

The beam combining instrument GRAVITY\footnote{http://www.mpe.mpg.de/2441478/Team}
 at the VLTI (P.I. Frank Eisenhauer, MPE, Garching)
will be able to measure the NIR image centroid paths during flux density excursions of SgrA*
(Eisenhauer et al. 2011, Straubmeier et al. 2012, Eckart et al. 2010, 2012).
These paths will depend on the geometrical structure and time evolution of the 
emitting region i.e. spot shape, e.g. presence of a torus or spiral-arm patterns, an emerging jet component.
Fig.~\ref{Fig:Gravity} shows the fringe tracking spectrometer of GRAVITY being under 
construction at the University of Cologne. Operating on six interferometric baselines, 
i.e. using all four UTs, the 2nd generation VLTI instrument GRAVITY will deliver narrow 
angle astrometry with 10$\mu$as accuracy at the infrared K-band.

While most of the mentioned geometries are currently able to fit the observed variable emission from SgrA*, 
future NIR interferometry with GRAVITY at the VLTI will break some of the degeneracies 
between different emission models. 
GRAVITY will be able to 
detect the positional shift of the photo-center of a flare at the Galactic Center 
within the $\sim$20~min orbital time scale of a source component close to the last stable orbit, 
using the flares as dynamical probes of the gravitational field around SgrA*. 
In particular GRAVITY observations of polarized NIR light could reveal a clear centroid 
track of bright spot(s) orbiting in the mid-plane of the accretion disk (e.g. Zamaninasab et al. 2010, 2011).
A non-detection of centroid shifts may point at a multi-component 
model or spiral arms scenarios. However, a clear wander between alternating centroid positions 
during the flares will strongly support the idea of bright long-lived spots occasionally orbiting 
the central black hole.

\vspace*{1cm}
\noindent
{\small
{\bf Acknowledgments:}
In this article we summarize contributions on the Galactic Center, held at 
the conference 'Equations of Motion in Relativistic Gravity' at the 
Physikzentrum Bad Honnef, Bad Honnef, Germany, (February 17-23, 2013)
\footnote{http://www.puetzfeld.org/eom2013.html}
and the COST action MP0905 - "Black Holes in a Violent Universe" - Galactic Center  
GC2013 conference\footnote{http://www.astro.uni-koeln.de/gc2013  and http://galacticcenter.iaa.es/}
titled 'The Galactic Center Black Hole Laboratory' held
at the Instituto de Astrofisica de Andalucia (IAA-CSIC) in Granada, Spain 
(November 19 - 22, 2013).
The main goal of the COST conference and the contribution in Bad Honnef 
was to focus on the new observational and
theoretical results that are linked to the Galactic Center and to point out
possible consequences for our understanding of galactic nuclei in general.
The COST conference brought together scientists working on all mass-scales of Black Holes 
(Quantum to super-massive), observers and theoreticians as well as particle physicists.

We are grateful to D. Merritt for constructive comments on the draft.
thank especially the Local Organizing Committees of the Bad Honnef and the Granada conferences.
We thank especially the Local Organizing Committees of the Bad Honnef and the Granada conferences.
For Bad Honnef we thank the organizers D. Puetzfeld, 
C. Laemmerzahl, B.F. Schutz as well as the 
staff of the Physikzentrum Bad Honnef, Bad Honnef, Germany.
The Granada conference was hosted by and held at the 
Instituto de Astrofisica de Andalucia (IAA-CSIC), Spain.
We thank especially the Local Organizing Committee at the IAA-CSIC that was lead by 
Dr. Antxon Alberdi (antxon@iaa.es).
The Granada conference was also closely linked to and substantially supported by the COST Action 
MP1104 - "Polarization as a tool to study the Solar System and beyond", the 
FP7 action "Probing Strong Gravity by Black Holes Across the Range of Masses"
[FP7-SPACE-2012-1 (FP7-312789)], and the 
collaborative research center SFB956 "Conditions and Impact of Star 
Formation - Astrophysics, Instrumentation and Laboratory Research".
We also thank RadioNet for support.
We are also grateful to all SOC members of the Granada conference:
Vladimir Karas, Michal Dovciak, Devaky Kunneriath, Delphine Porquet, Andreas Eckart, Silke Britzen,
Anton Zensus, Michael Kramer, Wolfgang Duschl, Antxon Alberdi , Rainer Sch\"odel, Murat H\"udaver,
Carole Mundell, Farhad Yusef-Zadeh, Andrea Ghez, Mark Morris.
This work was supported
in part by the Deutsche Forschungsgemeinschaft (DFG) via the Cologne Bonn
Graduate School (BCGS), the Max Planck Society through the International Max
Planck Research School (IMPRS) for Astronomy and Astrophysics, as well as
special funds through the University of Cologne. 
}

\vspace*{1cm}
\noindent
{\bf References:}
\rf{ Aharonian, F., et al.,"Pathway to the Square Kilometer Array", The German White Paper, 2013, arXiv:1301.4124}
\rf{ Akiyama, K.; Takahashi, R.; Honma, M.; Oyama, T.; Kobayashi, H.; 2013a, PASJ 65, 91}
\rf{ Akiyama, K.; Kino, M.; et al., 2013b, arXiv1311.5852A}
\rf{ Antonini \& Merritt 2013, ApJ 763 ,L10}
\rf{ Baganoff, F. K.; Maeda, Y.; Morris, M.; et al., 2003, ApJ 591, 891}
\rf{ Baganoff, F. K.; Bautz, M. W.; Brandt, W. N.; et al., 2001, Natur 413, 45}
\rf{ Ballone, A.; Schartmann, M.; Burkert, A.; et al., ApJ 776, 13}
\rf{ Barriere, N.M.; Tomsick, J.A.; Baganoff, F.K.; et al., 2014, arXiv1403.0900B}
\rf{ Bland-Hawthorn, J.; Maloney, Philip R.; Sutherland, Ralph S.; Madsen, G. J., 2013, ApJ 778, 58}
\rf{ Boller, Th.; M\"uller, A., 2013, Exciting Interdisciplinary Physics, FIAS Interdisciplinary Science Series. 
      ISBN 978-3-319-00046-6. Springer International Publishing Switzerland, 2013, p. 293}
\rf{ Bower, G.C., Falcke, H. , and  Backer, D.C, 1999, ApJ 523, L29}
\rf{ Bower, G.C. , Backer, D.C., et al., 1999, ApJ 521, 582}
\rf{ Bower, Geoffrey C., 2003, Ap\&SS 288, 69}
\rf{ Broderick, A.E.; Fish, V.L.; Doeleman, S.S.; Loeb, A.; 2011, ApJ 738, 38}
\rf{ Broderick, A.E.; Fish, V.L.; Doeleman, S.S.; Loeb, A.; 2011, ApJ 735, 110}
\rf{ Broderick, A.E.; Johannsen, T.; Loeb, A.; Psaltis, D.; 2014, ApJ 784, 7}
\rf{ Broderick, A.E.; Loeb, A., 2005, MNRAS 363, 353}
\rf{ Burkert, A.; Schartmann, M.; Alig, C.; et al., 2012, ApJ 750, 58}
\rf{ Chang, P., 2009, MNRAS, 393, 224}
\rf{ Chen, Xian; Amaro-Seoane, P., 2014, submitted (arXiv: 1401.6456)}
\rf{ Chandler, C.J.; Sjouwerman, L.O., The Astronomer's Telegram, No. 5727 and related ones}
\rf{ Church, R. P.; Tout, Christopher A.; Hurley, Jarrod R., 2009, PASA 26, 92}
\rf{ Combes, F.; Garcia-Burillo, S.; Casasola, V.; et al., 2014, arXiv1401.4120C}
\rf{ Crocker, R.M., 2012, MNRAS 423, 3512}
\rf{ Crocker, Roland M.; Aharonian, Felix, 2011, PhRvL 106, 1102}
\rf{ Crocker, R.M.; Jones, D.I.; Aharonian, F.; et al., 2011, MNRAS 411, L11}
\rf{ Crumley, P.; Kumar, P., 2013, MNRAS 436, 1955}
\rf{ Czerny, B., Kunneriath, D., Karas, V., Das, T.K., 2013, A\&A, 555, id.A97}
\rf{ Dale, J.; Davies, M.B.; Church, R.P.;  Freitag, M., 2009, MNRAS, 393, 1016}
\rf{ Dexter, J. \&  Fragile, P.Ch.,2013, MNRAS 432, 2252}
\rf{ Dexter, J.; Kelly, B.; Bower, G.C.; Marrone, D.P.; 2013, arXiv1308.5968D}
\rf{ Dexter, J.; Agol, E.; Fragile, P. C.; McKinney, J. C., 2012, JPhCS 372,id. 012023}
\rf{ Dexter, J., Agol, E., et al., 2010, ApJ, 717, 1092}
\rf{ Davies, M. B.; Church, R. P.; Malmberg, D.; Nzoke, S.; Dale, J.; Freitag, M., 2011, ASPC 439, 212}
\rf{ Eatough, R.P.; Falcke, H.; Karuppusamy, R.; 2013, Natur 501, 391}
\rf{ Eatough, R.; Karuppusamy, R.; Kramer, M.; et al., 2013, ATel 5027, 1}
\rf{ Eckart, A.; Muzic, K.; Yazici, S.; et al., 2013a, A\&A 551, 18}
\rf{ Eckart, A.; Britzen, S.; Horrobin, M.; et al., 2013b, arXiv1311.2743E}
\rf{ Eckart, A.; Horrobin, M.; Britzen, S.; et al., 2013c, arXiv1311.2753E}
\rf{ Eckart, A.; Garcia-Marin, M.; et al., 2012, A\&A 537, 52}
\rf{ Eckart, A., Eckart, A.; Garcia-Marin, M.; Vogel, S. N.; Teuben, P., et al. 2012 A\&A 537, 52}
\rf{ Eckart, A.; Sabha, N.; Witzel, G.; Straubmeier, C., 2012, SPIE 8445, 1}
\rf{ Eckart, A.; Zamaninasab, M.; Straubmeier, C.; et al. 2010, SPIE 7734, 27}
\rf{ Eckart, A.; Baganoff, F. K.; Zamaninasab, M.; Morris, M. R.; Sch\"odel, R.; et al., 2008a, A\&A 479, 625}
\rf{ Eckart, A.; Sch\"odel, R.; Garcia-Marin, M.; Witzel, G.; Weiss, A.; et al., 2008b, A\&A 492, 337}
\rf{ Eckart, A.; Sch\"odel, R.; Meyer, L.; Trippe, S.; Ott, T.; Genzel, R., 2006a, A\&A 455, 1}
\rf{ Eckart, A.; Baganoff, F. K.; Sch\"odel, R.; Morris, M.; et al, 2006b, A\&A 450, 535}
\rf{ Eckart, A.; Genzel, R., 1997, MNRAS 284, 576}
\rf{ Eisenhauer, F.; Perrin, G.; Brandner, W.; Straubmeier, C.; et al., 2011, Msngr 143, 16}
\rf{ Eisenhauer, F.; Sch\"odel, R.; Genzel, R.; Ott, T.; Tecza, M.; Abuter, R.; Eckart, A.; Alexander, T., 2003, ApJ 597, L121}
\rf{ Eisenhauer, Frank, 2010, IAUS 261, 269}
\rf{ Eilon, E., Kupi, G., \& Alexander, T. 2009, ApJ, 698, 641}
\rf{ Falcke, H.; Markoff, S. B.,  2013, Classical and Quantum Gravity, 30, Issue 24, id. 244003}
\rf{ Falcke, H., \& Markoff, S. 2000, A\&A, 362, 113}
\rf{ Feldmeier, A.; L\"utzgendorf, N.; Neumayer, N.; et al.,  2013, A\&A 554, 63}
\rf{ Fish, V.L.; Doeleman, S.; Krichbaum, T.; Zensus, A.; Event Horizon Telescope Collaboration;; 2014, AAS 22344304}
\rf{ Fish, V.L.; Doeleman, S.S.; Beaudoin, C.; Blundell, R.; Bolin, D.E.; Bower, G.C.; et al.; 2011, ApJ 727, L36}
\rf{ Frank, J.; Rees, M. J.; 1976, MNRAS 176, 633}
\rf{ Garcia-Burillo, S.; Combes, F., 2012, JPhCS, 372, id 2050}
\rf{ Garcia-Burillo, S.; Usero, A.; Alonso-Herrero, A.; et al., 2012, A\&A 539, 8}
\rf{ Garcia-Burillo, S.; Combes, F.; Hunt, L. K.; et al., 2003, A\&A 407, 485}
\rf{ Ghez, A.M., Witzel, G., Sitarski B., Meyer L., et al.; 2014, ATel \#6110; (http://www.astronomerstelegram.org/?read=6110)}
\rf{ Ghez, A. et al., 2009, Astro2010: The Astronomy and Astrophysics Decadal Survey, Science White Papers, no. 89}
\rf{ Ghez, A. M.; Salim, S.; Hornstein, S. D.; et al., 2005, ApJ 620, 744}
\rf{ Gillessen, S.; Genzel, R.; Fritz, T. K.; et al., 2012, Natur 481, 51}
\rf{ Gillessen, S., Eisenhauer, F., Trippe, S., et al. 2009a, ApJ, 692, 1075}
\rf{ Gillessen, S.; Eisenhauer, F.; Fritz, T.K.; et al., 2009b, ApJ 707, L114}
\rf{ Gould R.J., 1979, A\&A 76, 306}
\rf{ Haas, J.; Subr, L.; Vokrouhlicky, D., 2011, MNRAS 416, 1023}
\rf{ Haas, J.; Subr, L., 2012, JPhCS.372, id. 2059}
\rf{ Heckman, T.; Best, P., 2014arXiv1403.4620H}
\rf{ Hopman, C. \& Alexander, T. 2006a, ApJ, 645, 1152}
\rf{ Huang, L.; Shen, Z.-Q.; Gao, F.	; 2012, ApJ 745, L20}
\rf{ Huang, Lei; Cai, M.; Shen, Z.-Q.; Yuan, F., 2007, MNRAS 379, 833}
\rf{ Jalali, B.; Pelupessy, I.; Eckart, A.; 2013, arXiv1311.4881J}
\rf{ Karas, V.; Dovciak, M.; Zamaninasab, M.; Eckart, A., 2011, ASPC 439, 344}
\rf{ Karas, V.; Kopacek, O.; Kunneriath, D., 2012, Classical and Quantum Gravity, 29, id. 035010}
\rf{ Karas, V.; Kopacek, O.; Kunneriath, D., 2013, International Journal of Astronomy and Astrophysics, 3, 18}
\rf{ Koide, S.; Arai, K., 2008, ApJ, 682, 1124}
\rf{ Krips, M.; Eckart, A.; Krichbaum, T. P.; et al., 2007, A\&A 464, 553}
\rf{ Lee, K.J.; Eatough, R.; Karuppusamy, R.; 2013, ATel 5064, 1}
\rf{ Lu, R.-S.; Krichbaum, T.P.; Eckart, A.; K\"onig, S.; Kunneriath, D.; Witzel, G.; Witzel, A.; Zensus, J. A.; 2011, A\&A 525, 76}
\rf{ Macquart, J.-P.,  Bower, G.C.,  Wright, M.C.H. , et al., 2006, ApJ 646, L111}
\rf{ Marquez, I.; Masegosa, J., 2008, RMxAC 32, 150}
\rf{ Marscher, A.P., 1983, ApJ 264, 296}
\rf{ Merritt, D. 2012, Dynamics and Evolution of Galactic Nuclei", (Princeton University Press)}
\rf{ Merritt, D., Alexander, T., Mikkola, S., \& Will, C.~M., 2010, PhRvD 81, 062002}
\rf{ Merritt, D., Alexander, T., Mikkola, S., \& Will, C.~M., 2011, PhRvD 84, 044024}
\rf{ Marrone, D. P., Baganoff, F. K., Morris, M., et al. 2008, ApJ, 682, 373}
\rf{ Mastrobuono-Battisti, Alessandra; Perets, Hagai B.,  2013, ApJ 779, 85}
\rf{ Mouawad, N.; Eckart, A.; Pfalzner, S.; Sch\"odel, R.; Moultaka, J.; Spurzem, R.; 2005, AN326, 83}
\rf{ Mouawad, N.; Eckart, A.; Pfalzner, S.; Moultaka, J.; Straubmeier, C.; Spurzem, R.; Sch\"odel, R.; Ott, T.; 2003, ANS 324, 315}
\rf{ Moser, L.; Zuther, J.; Fischer, S.; Busch, G.; et al., 2013, arXiv1309.6921M}
\rf{ Mundell, C.G.; James, P.A.; Loiseau, N.; et al., 2004, ApJ 614, 648}
\rf{ Mundell, C. G.; Wrobel, J. M.; Pedlar, A.; 2003, ApJ 583, 192}
\rf{ Mundell, C. G.; Dumas, G.; et al., 2007, NewAR 51, 34}
\rf{ Mundell, C.G.; Ferruit, P.; Nagar, N.; Wilson, A. S., 2009, ApJ 703,802}
\rf{ Mori, K.; Gotthelf, E.V.; Zhang, S.; et al., 2013, ApJ 770, L23}
\rf{ Morozova, V.S.; Rezzolla, L.; Ahmedov, B.J., 2014, submitted (arXiv:1310.3575)}
\rf{ Moscibrodzka, M.; Falcke, H., 2013, A\&A 559, L3}
\rf{ Meyer, Leo; Ghez, A. M.; Do, T.; Boehle, A.; et al.  2014, AAS22310807M}
\rf{ Meyer, L.; Ghez, A. M.; Witzel, G.; et al., 2013, arXiv1312.1715M	}
\rf{ Meyer, L.; Ghez, A. M.; Sch\"odel, R.; Yelda, S.; Boehle, A.; et al.; 2012, Sci 338, 84}
\rf{ Muzic, K.; Eckart, A.; Sch\"odel, R.; et al. 2010, A\&A 521, 13}
\rf{ Narayan, R.; \"Ozel, F.; Sironi, L., 2012, ApJ 757, L20}
\rf{ Narayan, R.; Mahadevan, R.; Grindlay, J. E.; Popham, R. G.; Gammie, C., 1998, ApJ 492, 554}
\rf{ Nayakshin, S.; Cuadra, J.; Springel, V., 2007, MNRAS 379, 21}
\rf{ Nowak, M. A.; Neilsen, J.; Markoff, S. B.; et al., 2012, ApJ 759, 95}
\rf{ Perets, Hagai B.; Mastrobuono-Battisti, Alessandra, 2014, arXiv1401.1824P}
\rf{ Phifer, K.; Do, T.; Meyer, L.; et al., 2013, ApJ 773, L13}
\rf{ Porquet, D.; Grosso, N.; Predehl, P.; et al., 2008, A\&A 488, 549}
\rf{ Porquet, D.; Predehl, P.; Aschenbach, B.; Grosso, N.; et al., 2003, A\&A 407, L17}
\rf{ Psaltis, D., 2012, ApJ, 759, 130}
\rf{ Rauch, K. P. \& Tremaine, S. 1996, New A, 1, 149}
\rf{ Rea, N.; Esposito, P.; Pons, J. A.; Turolla, R.; et al., 2013, ApJ 775, L34}
\rf{ Rubilar, G. F.; Eckart, A., 2001, A\&A 374, 95}
\rf{ Sabha, N., Merritt, D.; et al., 2012, A\&A 545, 70}
\rf{ Sadowski, A.; Sironi, L.; Abarca, D.; et al., 2013, MNRAS 432, 478}
\rf{ Schartmann, M.; Burkert, A.; Alig, C.; et al., 2012, ApJ 755, 155}
\rf{ Scoville, N.; Burkert, A., 2013, ApJ 768, 108}
\rf{ Shannon, R.M.; Johnston, S., 2013, MNRAS 435, L29}
\rf{ Shcherbakov \& Baganoff (2010) }
\rf{ Shcherbakov, R.V., 2014, ApJ 783, 31}
\rf{ Spitler, L.G.; Lee, K.J.; Eatough, R.P.; Kramer, M.; 2014, ApJ 780, L3}
\rf{ Soffitta, P.; Campana, R.; Costa, E.; et al.  2012, SPIE 8443, 1}
\rf{ Soffitta, P.; Barcons, X.; Bellazzini, R.; et al., 2013, ExA 36, 523}
\rf{ Straubmeier, C.; Fischer, S.; Araujo-Hauck, C.; et al., 2012, SPIE 8445, 2}
\rf{ Su, M.; Slatyer, T.R.; Finkbeiner, D.P., 2010, ApJ 724, 1044}
\rf{ Subr, L.; Karas, V.; Hure, J.-M., 2004, MNRAS, 354, 1177}
\rf{ Subr, L.; Karas, V., 2005, A\&A 433, 405}
\rf{ Subr, L.; Haas, J., 2012, JPhCS 372, id. 2018}
\rf{ Valencia-S, M.; Bursa, M.; Karssen, G.; Dovciak, M.; Eckart, A.; Horak, J.; Karas, V., 2012, JPhCS, 372, 2073}
\rf{ Will, C.~M., 2008, ApJ 674, L25}
\rf{ Witzel et al. 2012 ApJS 203, 18 }
\rf{ Yoshikawa, T.; Nishiyama, S.; et al., 2013, ApJ 778, 92}
\rf{ Yuan, Feng; Shen, Zhi-Qiang; Huang, Lei, 2006, ApJ 642, L45}
\rf{ Yusef-Zadeh, F.; Wardle, M., 2013, ApJ 770, L21}
\rf{ Yusef-Zadeh, F.; Royster, M.; Wardle, M.; et al., 2013, ApJ 767, L32}
\rf{ Yusef-Zadeh, F.; Hewitt, J. W.; Arendt, R. G.; et al., 2009, ApJ 702, 178}
\rf{ Yusef-Zadeh, F.; Lacy, J. H.; Wardle, M.; et al., 2010, ApJ 725, 1429}
\rf{ Yusef-Zadeh, F.; Roberts, D.; Wardle, M.; et al.,  2006, ApJ 650, 189}
\rf{ Zamaninasab, M.; Eckart, A.; Dovciak, M.; Karas, V.; et al., 2011, MNRAS 413, 322}
\rf{ Zamaninasab, M.; Witzel, G.; Eckart, A.; Sabha, N.; et al., 2011, ASPC 439, 323}
\rf{ Zamaninasab, M.; Eckart, A.; Witzel, G.; Dovciak, M.; Karas, V.; et al., 2010, A\&A 510, 3}
\rf{ Zajacek, M, Karas, V.,  Eckart, A., accepted by A\&A 2014}
\rf{ Zubovas, K. \& Nayakshin, S., 2012, MNRAS 424, 666}
\rf{ Zucker, S.; Alexander, T.; et al., 2006, ApJ 639, L21}
\end{document}